\documentclass[aps,pre,amsmath,amssymb,twocolumn]{revtex4}
\usepackage{amsmath,amsthm,amssymb}
\usepackage{graphicx}
\usepackage{pstricks}

\begin{document}

\title {Transparent Quantum Graphs}
\author{J.R. Yusupov$^1$, K.K. Sabirov$^{2}$, M. Ehrhardt$^{3}$  and D.U. Matrasulov$^1$}
\affiliation{$^1$Turin Polytechnic University in Tashkent, 17
Niyazov Str., 100095, Tashkent, Uzbekistan\\
$^2$Tashkent University of Information Technologies, 108 Amir
Temur Str., 100200, Tashkent Uzbekistan\\
$^3$Bergische Universit\"at Wuppertal, Gau{\ss}strasse 20, D-42119 Wuppertal,
Germany}

%%%%%%%%%%%%%%%%%%%%%%%%%%%%%%%%%%%%%%
\begin{abstract}
We consider quantum graphs with transparent branching points. To design such
networks, the concept of transparent boundary conditions is applied to the
derivation of the vertex boundary conditions for the linear Schr\"odinger
equation on metric graphs. This allows to derive simple constraints, which use
equivalent usual Kirchhoff-type boundary conditions at the vertex to the
transparent ones. The approach is applied to quantum star and tree graphs.
However, extension to more complicated graph topologies is rather straight
forward.
\end{abstract}
\maketitle

%%%%%%%%%%%%%%%%%%%%%%%%%%%%%%%%%%%%%%
\section{Introduction}

The problem of wave transport in branched structures and networks is of
importance for many areas of contemporary physics, such as optics, fluid
dynamics, condensed matter and polymers. Optical and quantum mechanical waves
propagation in such systems appears e.g., in different branched waveguides
\cite{BWG1,BWG2,BWG3,BWG4,BWG5,BWG6,BWG7}. Effective transfer of light, charge, heat,
spin and signal in networks requires solving the problem of tunable wave
dynamics. This can be achieved using the models, which provide realistic
description of particle and wave transport in branched structures.

An important feature of particle and wave dynamics in networks is the
transmission through the branching points, which is usually accompanied by the
reflection (backscattering) of a wave at these points. Dominating of reflection
compared to transmission implies large ``resistivity'' of a network with
respect to the wave propagation. Therefore, it is important from the viewpoint
of practical applications, to reduce such resistivity by providing a minimum of
reflection, or by its absence. This task leads to the problem of tunable
transport in branched structures, whose ideal result should be reflectionless
transmission of the waves through the branching points of the structure. From
the viewpoints of practical applications in condensed matter, such transmission
implies ballistic transport of charge, spin, heat and other carriers in
low-dimensional branched materials. The latter is of importance for the
effective utilization of different functional materials.

Reflectionless transport of waves in optical fiber networks is one of the
central tasks for the fiber optics, as many information-communication devices
(e.g., computers, computer networks, telephones, etc.) use optical fiber
networks for information (signal) transfer. Such networks are also used in
different optoelectronic devices. High speed and lossless transfer of
information in such devices require minimum of backscattering or its absence.
Important areas, where the reflectionless, or ballistic transport in branched
structures is required, are molecular electronics and conducting polymers. Both
deal with the modeling of reflectionless wave transport in networks and require
revealing the conditions providing such regime. Different quantum wire networks
appearing in solid state physics are also potential structures, where the wave
transport can be tuned from diffusive to ballistic regime using the transparent
boundary conditions.

In this paper we study the problem of reflectionless transport in quantum
networks, which are modeled in terms of the linear Schr\"odinger equations on
metric graphs. To describe the regime, when there is no backscattering at the
network branching points, we use the concept of transparent boundary conditions
for the time-dependent Schr\"odinger equation studied earlier in detail in the
Refs.~\cite{Arnold1998,Ehrhardt1999,Ehrhardt2001,Ehrhardt2002,Arnold2003,Jiang2004,Antoine2008,Ehrhardt2008,Sumichrast2009,Antoine2009,Ehrhardt2010,Klein2011,Arnold2012,Feshchenko2013,Antoine2014,Petrov2016}.
Combining transparent boundary conditions concept with that of vertex boundary
conditions for the linear Schr\"odinger equation on metric graphs, we derive
the constraints, providing the regime of reflectionless transmission of the
waves through the graph branching points (vertices).

More precisely, we reveal the condition for the regime, at which Kirchhoff-type
boundary conditions at the branching point become equivalent to the transparent
boundary conditions. The condition is obtained in a simple form of the sum rule
for the coefficients, which can characterize physical properties of the network
branches. We note that the linear and nonlinear wave equations on metric graphs
have been studied earlier in the context of quantum graphs
\cite{Uzy1,Hul,Kuchment04,Uzy2,Exner15} and soliton dynamics in networks
\cite{Zarif,zar2011,caputo14,Our2015,Adami16,Our1,Adami17,dimarecent,Karim2018,KarimNLDE,Our2018}.
It was found in the Refs.~\cite{Zarif,Our1,KarimNLDE} that transmission of the
waves through the network branching point can be reflectionless, when provided
certain constraints are fulfilled. However, no strict explanation for such
effect have been presented in those studies.

This paper can be considered as the first step in the way for the formulation
of clear, physically reasonable, strict mathematical conditions for the
reflectionless wave transmission in network branching points in quantum regime.
The paper is organized as follows. In the next section we briefly recall the
concept of transparent boundary conditions for the linear Schr\"odinger
equation on a line. In Section III we recall the concept of quantum graphs.
Section IV provides extension of the concept of transparent boundary conditions
to quantum networks described by the linear Schr\"odinger equation on metric
graphs and presents some numerical results. Finally, Section V presents some
concluding remarks.

%%%%%%%%%%%%%%%%%%%%%%%%%%%%%%%%%%%%%%%%%%
\section{Transparent boundary conditions in quantum mechanics}

The problem of transparent boundary conditions (TBC) for the wave
equations has attracted much attention in different practically
important contexts (see, e.g., papers
\cite{Arnold1998,Ehrhardt1999,Ehrhardt2001,Ehrhardt2002,Arnold2003,Jiang2004,Antoine2008,Ehrhardt2008,Sumichrast2009,Antoine2009,Ehrhardt2010,Klein2011,Arnold2012,Feshchenko2013,Antoine2014,Petrov2016}
for review). Such boundary conditions describe the absorption or
reflectionless transmission of particles (waves) at the boundary
of two domains. Approximate TBCs for the linear Schr\"odinger
equation were formulated by Shibata \cite{Shibata} and Kuska
\cite{Kuska}, where dispersion relations for the plane waves is
approximated to derive TBCs. A strict mathematical analysis of TBC
for different wave equations, including quantum mechanical
Schr\"{o}dinger equation can be found in the
Refs.~\cite{Ehrhardt2002,Arnold2003,Ehrhardt2008,Antoine2014,Petrov2016}.

The explicit form of such boundary conditions are much more complicated than
those of Dirichlet, Neumann and Robin conditions. Therefore, their
discretization causes a serious accuracy loss in the numerical computations and
also often reduce the stability range of the overall scheme. Here we briefly
recall the formulation of TBCs on a line following the
Refs.~\cite{Ehrhardt1999,Ehrhardt2001,Ehrhardt2002,Arnold2003}.

Let us consider the wave (particle) motion in a 1D domain $(-\infty,\; +\infty)$ and
described by the following time-dependent Schr\"odinger equation (with
$\hbar=m=1$):
\begin{equation}
i\frac{\partial}{\partial t} \Psi=-\frac12\frac{\partial^2}{\partial x^2} \Psi
+ V(x,t)\Psi,\label{lse03}
\end{equation}
with the initial condition given as
\begin{equation*}
\Psi(x,0)=\Psi^I(x),
\end{equation*}
where $\Psi^I\in L^2(\mathbb{R})$, $V(.,t)\in L^\infty(\mathbb{R})$ and
$V(x,t)$ is an external potential. Our purpose is to formulate the boundary
conditions for Eq.~\eqref{lse03}, which provide reflectionless transmission of
the wave via the given points, $0$ and $L$. One of the prescriptions to design
such boundary conditions was developed in the
Refs.~\cite{Ehrhardt1999,Ehrhardt2001,Ehrhardt2002} and uses the following two
basic assumptions: the initial data $\Psi^I$ is compactly supported in the
computational domain $0 < x < L$, and the given external potential is constant
outside this finite domain, i.e.\ $V(x,t)=0$ for $x\leqslant 0, V(x,t)=V_L$ for
$x\geqslant L$.

Then we can separate the whole problem into the so-called ``interior'' and
``exterior'' problems. The ``interior'' problem is given by
\begin{align}
&i\frac{\partial}{\partial t} \Psi=-\dfrac12\frac{\partial^2}{\partial x^2} \Psi+V(x,t)\Psi, \ 0<x<L,\ t>0,\label{intp1}\\
&\Psi(x,0)=\Psi^I(x),\nonumber\\
&\frac{\partial}{\partial x}\Psi(0,t)=(T_0\Psi)(0,t),\nonumber\\
&\frac{\partial}{\partial x}\Psi(L,t)=(T_L\Psi)(L,t),\nonumber
\end{align}
where $T_{0,L}$ denotes the Dirichlet-to-Neumann (DtN) maps at the boundaries, which
can be  obtained by solving the two ``exterior'' problems given as
\begin{align}
&i\frac{\partial}{\partial t} v=-\dfrac12\frac{\partial^2}{\partial x^2} v+V_Lv, \ x>L,\ t>0,\label{extp1}\\
&v(x,0)=0,\nonumber\\
&v(L,t)=\Phi(t),\ \ \ \ t>0, \ \ \Phi(0)=0,\nonumber\\
&v(\infty,t)=0,\nonumber\\
&(T_L\Phi)(t)=\frac{\partial}{\partial x}v(L,t),\nonumber
\end{align}
and it can analogously be done for $T_0$.

Using the Laplace transformation
\begin{equation}
\hat v(x,s)=\int\limits_0^\infty{v(x,t)e^{-st}dt},\label{laplace}
\end{equation}
the right ``exterior'' problem is transformed to
\begin{align}
&\frac{\partial^2}{\partial x^2} \hat v+2i(s+iV_L)\hat v=0, \ \ x>L,\label{extp2}\\
&\hat v(L,s)=\hat\Phi(s).\nonumber
\end{align}

The general solution for Eq.~\eqref{extp2} can be written as
\begin{equation*}
\hat
v(x,s)=C_1e^{\sqrt[+]{-2i(s+iV_L)}(x-L)}+C_2e^{-\sqrt[+]{-2i(s+iV_L)}(x-L)}.
\end{equation*}
Since its solutions have to decrease as $x\to +\infty$, we obtain
\begin{equation}
\hat v(x,s)=e^{-\sqrt[+]{-2i(s+iV_L)}(x-L)}\hat\Phi(s),
\end{equation}
where $\sqrt[+]{\ \ }$ denotes the square root with nonnegative real part.

Hence the Laplace-transformed Dirichlet-to-Neumann operator $T_L$
reads
\begin{equation}
\widehat{T_L\Phi}(s)=\frac{\partial}{\partial x}\hat
v(L,s)=-\sqrt{2}e^{-i\pi/4}\sqrt[+]{s+iV_L}\hat\Phi(s),
\end{equation}
and $T_0$ is calculated analogously.

An inverse Laplace transformation yields the right TBC at $x=L$:
\begin{equation}
\frac{\partial}{\partial
x}\Psi(x=L,t)=-\sqrt{\frac{2}{\pi}}e^{-i\frac{\pi}{4}}e^{-iV_Lt}\frac{d}{dt}\int\limits_0^t{\frac{\Psi(L,\tau)e^{iV_L\tau}}{\sqrt{t-\tau}}d\tau}.
\end{equation}

Similarly, the left TBC at $x = 0$ is obtained as
\begin{equation}
\frac{\partial}{\partial
x}\Psi(x=0,t)=\sqrt{\frac{2}{\pi}}e^{-i\frac{\pi}{4}}\frac{d}{dt}\int\limits_0^t{\frac{\Psi(0,\tau)}{\sqrt{t-\tau}}d\tau}.\label{tbc00}
\end{equation}

In this paper we apply this procedure to quantum graphs. It should be noted
that no practical applications of the TBCs in physical systems have been
considered so far. Here we will do this for quantum networks, which appear in
different branches of optics, condensed matter and polymer physics.

%%%%%%%%%%%%%%%%%%%%%%%%%%%%%%%%%%%%%%%%%%%%%%%%%%%%%
\section{Quantum graphs}

The concept of quantum graphs has been introduced first by Exner, Seba and
Stovicek to describe free quantum motion on branched wires \cite{Exner1}.
However, the pioneering treatment of the quantum mechanical motion in branched
structures dates back to the Refs.~\cite{Pauling,Rud,Alex}.

Later Kostrykin and Schrader derived the general boundary conditions providing
self-adjointness of the Schr\"odinger operator on graphs \cite{Kost}.
Relativistic quantum mechanics described by Dirac \cite{Bolte} and
Bogoliubov-de Gennes operators \cite{KarimBdG} on graphs have been studied
recently. Hul \emph{et al} considered experimental realization of quantum
graphs in optical microwave networks \cite{Hul}. An important topic related to
quantum graphs was studied in the context of quantum chaos theory and spectral
statistics \cite{Uzy1,Bolte,Uzy2}. Spectral properties and band structure of
periodic quantum graphs also attracted much interest \cite{Exner15,Grisha}.
Different aspects of the Schr\"odinger operators on graphs have been studied in
the Refs.~\cite{Exner15,Grisha,Mugnolo}. Tunable directed transport in quantum
driven graphs has been also studied in \cite{yusupov2015}, where the reflection
at the branching point has been discussed.

Nonlinear extensions of the quantum graph concept have been studied in the
context of soliton transport in networks in the
Refs.~\cite{Zarif,zar2011,caputo14,Our2015,Our1,KarimNLDE,Our2018}. In
\cite{Zarif} the nonlinear Schr\"odinger equation on metric graphs is studied
and condition for integrability is derived in the form of a sum rule for
nonlinearity coefficients. In \cite{zar2011} such study is extended to
Ablowitz-Laddik equation. The stationary Schr\"odinger equation on metric
graphs and standing wave solitons in networks are studied in
\cite{Karim2013,Adami16,Adami17}. In \cite{Ourhe} the linear and nonlinear head
equations on metric star graphs is studied in the context of thermal diffusion.
The nonlinear Schr\"odinger equation with the subcritical power nonlinearity on
a metric star graph generalized Kirchhoff boundary conditions in
\cite{dimarecent}, where the stability of half-soliton solutions is also
analyzed. Integrable sine-Gordon equation on metric graphs is studied in
\cite{caputo14,Our1,Karim2018}. Linear and nonlinear systems of PDE on metric
graphs are considered in \cite{Bolte,KarimBdG,KarimNLDE}.

Quantum graphs, which are the one- or quasi-one dimensional branched quantum
wires, can be modeled in terms of quantum mechanical wave equations on metric
graphs by imposing the boundary conditions at the branching points (vertices)
and bond ends. The metric graphs are the set of bonds with assigned length and
which are connected to each other at the vertices. The connection rule is
called topology of a graph and given in terms of the adjacency matrix
\cite{Uzy1,Uzy2}:
\begin{equation}
C_{ij}=C_{ji}=\begin{cases}1 & \text{ if }\; i\;
\text{ and }\; j\; \text{ are connected, }\\
 0 & \text{ otherwise, }\end{cases}
\end{equation}
for $i,j=1,2,\dots,N$.

\begin{figure}[th!]
\includegraphics[width=70mm]{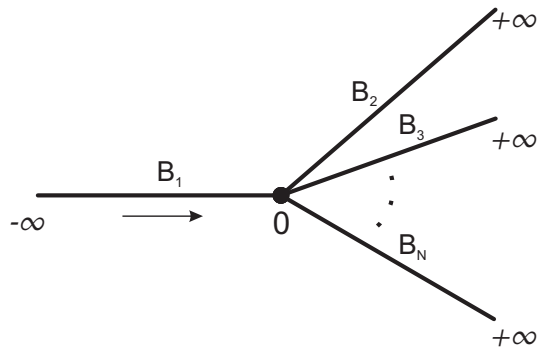}
\caption{Sketch of a star graph with $N$ semi-infinite bonds.} \label{pic1}
\end{figure}

The graph with the simplest topology is called star graph. It has one branching
point, which connects three or more bonds (see, Fig.~\ref{pic1}). The particle
(wave) dynamics in a quantum graph is described in terms of the linear
Schr\"odinger equation (with $\hbar=m=1$), which can be written for each bond
of a star graph with $N$ bonds as
\begin{equation}
i\frac{\partial\Psi_j}{\partial t} =
\left[-\frac12\frac{d^2}{dx^2}+V_j(x,t)\right]\Psi_j,\;\; j=1,2,\dots,N, \label{se01}
\end{equation}
where $\Psi_j =\Psi_j(x,t)$ is the wave function of $j$th bond, $V_j(x,t)$ is
the potential acting on $j$th bond and $j$ is the bond number. In the
following, for simplicity we assume that $V_j=0$. To solve Eq.~\eqref{se01} one
needs to impose initial condition and boundary conditions at the branching
point and bond ends. Most general boundary conditions providing
self-adjointness of the Schr\"odinger operator on graphs have been derived in
\cite{Kost}. Physically reasonable vertex boundary conditions, which are
consistent with these general ones can be chosen as the continuity of the wave
function at the vertex and Kirchhoff's rule \cite{Uzy1,Uzy2}:
\begin{equation}
\Psi_1(0,t)= \Psi_2(0,t) =\dots= \Psi_N(0,t),
\end{equation}
\begin{equation}
\frac{\partial}{\partial
x}\Psi_1(x=0,t)-\sum_{j=2}^{N}\frac{\partial}{\partial
x}\Psi_j(x=0,t) = 0.
\end{equation}

According to the scattering approach \cite{Uzy1} this ``natural'' boundary
conditions, given at a vertex $i$, cause reflection of a wave packet with
back-scattering amplitude $s_j=-1+2/N$ and the reflection probability $|s_j|^2$
approaches 1 as the number of bonds increases. Other details of the solution of
the stationary version of Eq.~\eqref{se01} can be found elsewhere (see, e.g.\
\cite{Uzy1,Uzy2}). Here we will focus on the dynamical problem, which is
described in terms of the time-dependent Schr\"odinger equation, by considering
the wave transport in quantum graphs.

%%%%%%%%%%%%%%%%%%%%%%%%%%%%%%%%%%%
\section{Transparent quantum graphs}

Here we combine the above concepts of transparent boundary conditions and
quantum graphs to design the vertex boundary conditions providing
reflectionless transmission of waves through the graph branching points. In the
following such graphs will be called ``transparent quantum graphs''. Using such
approach, below we show that Kirchhoff-type boundary conditions (weight
continuity and current conservation) can become equivalent to transparent
vertex boundary conditions under constraints in the form of a simple sum rule
that is fulfilled for the weight coefficients. For simplicity (without losing
generality) we consider first a star graph with three bonds. We assign to each
bond $B_j$ a coordinate $x_j$, which indicates the position along the bond: for
bond $B_1$ it is $x_1\in (-\infty,0]$ and for $B_{1,2}$ they are $x_{2,3}\in
[0,+\infty)$. In the following we use the shorthand notation $\Psi_j(x)$ for
$\Psi_j(x_j)$ and it is understood that $x$ is the coordinate on the bond $j$
to which the component $\Psi_j$ refers. Then for the wave dynamics in such
graph one can write the time-dependent Schr\"odinger equation given by (in the
units $\hbar=m=1$)
\begin{equation}
i\frac{\partial}{\partial t} \Psi_j(x,t)=-\frac12\frac{\partial^2}{\partial
x^2} \Psi_j(x,t), \ \ \ \ j=1,2,3\label{lse0001}
\end{equation}
At the branching point (vertex) we impose the boundary conditions
in the form of the wave function weight continuity given by
\begin{equation}
\alpha_1\Psi_1(0,t)=\alpha_2\Psi_2(0,t)=\alpha_3\Psi_3(0,t),\label{cont1}
\end{equation}
and Kirchhoff type conditions, which is given as
\begin{align}
\frac{1}{\alpha_1}\frac{\partial}{\partial
x}\Psi_1(x=0,t)&=\frac{1}{\alpha_2}\frac{\partial}{\partial x}
\Psi_2(x=0,t)\nonumber\\
&+\frac{1}{\alpha_3}\frac{\partial}{\partial x}
\Psi_3(x=0,t).\label{Kirch1}
\end{align}

In the following we consider the wave going from the first to second and third
bonds, i.e., the initial condition is compactly supported in the first bond.
Then the ``interior'' problem for the first bond, $B_1$ can be written as
\begin{align}
&i\frac{\partial}{\partial t} \Psi_1=-\dfrac12\frac{\partial^2}{\partial x^2} \Psi_1, \ \ x<0,\ \ t>0,\\
&\Psi_1(x,0)=\Psi^I(x),\nonumber\\
&\frac{\partial}{\partial x}\Psi_1(0,t)=(T_0\Psi_1)(0,t).\nonumber
\end{align}

Corresponding ``exterior'' problems for $B_{2,3}$ are given by
\begin{align}
&i\frac{\partial}{\partial t} \Psi_{2,3}=-\dfrac12\frac{\partial^2}{\partial x^2} \Psi_{2,3}, \ \ x>0,\ \ t>0,\\
&\Psi_{2,3}(x,0)=0,\nonumber\\
&\Psi_{2,3}(0,t)=\Phi(t),\ \ \ \ t>0, \ \ \Phi(0)=0,\nonumber\\
&(T_0\Phi)(t)=\frac{\partial}{\partial x}\Psi_{2,3}(0,t).\nonumber
\end{align}

\begin{figure}[th!]
\includegraphics[width=84mm]{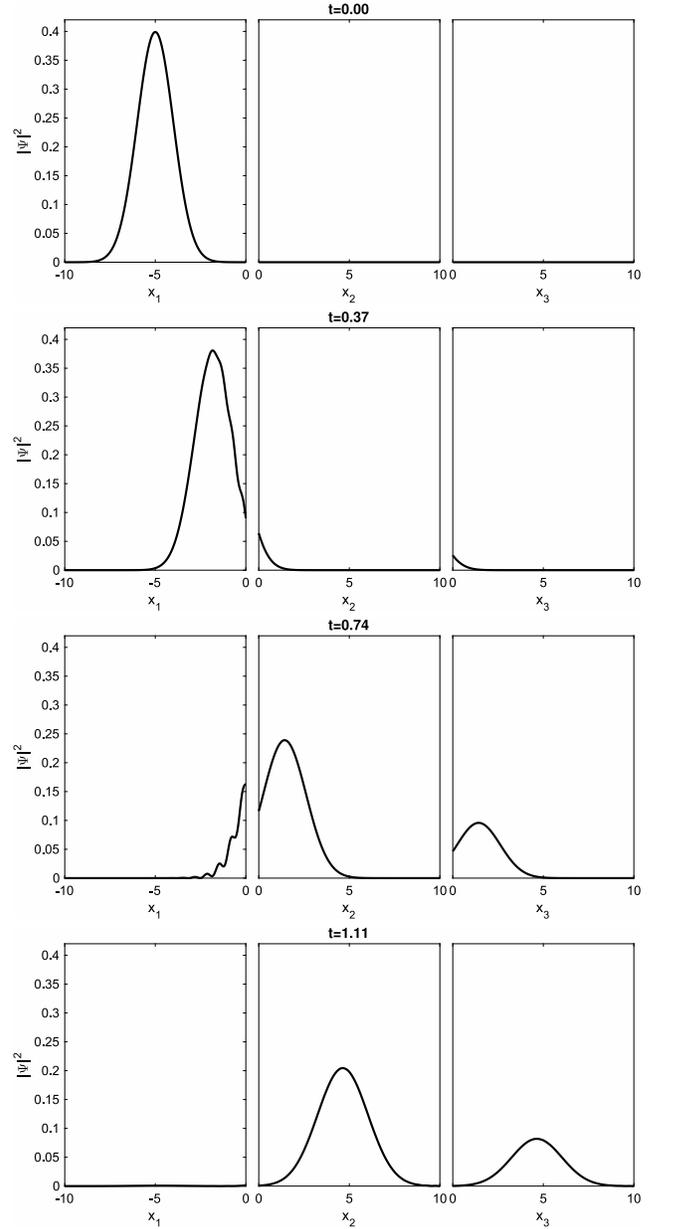}
\caption{The profile of the wave function plotted at different time moments for
the regime when the sum rule is fulfilled (no reflection is occurred):
$\alpha_1=1/\sqrt{1/20+1/50}$, $\alpha_2=\sqrt{20}$
and $\alpha_3=\sqrt{50}$.
Each column number (from the left to the right) corresponds to a bond number.}
\label{fig2}
\end{figure}

Finally, using the Laplace transformation \eqref{laplace}, the two ``exterior''
problems are transformed to
\begin{align}
&\frac{\partial^2}{\partial x^2} \hat\Psi_{2,3}+2is\hat\Psi_{2,3}=0, \ \ \ \ x>0,\\
&\hat\Psi_{2,3}(0,s)=\hat\Phi(s).\nonumber\\
\end{align}

The general solution of the exterior problems can be written as
\begin{equation}
\hat\Psi_{2,3}(x,s)=C_1e^{\sqrt[+]{-2is}x}+C_2e^{-\sqrt[+]{-2is}x}.
\end{equation}
Since we have $\hat\Psi_{2,3}\in L^2(0,\infty)$, we obtain
\begin{equation}
\hat\Psi_{2,3}(x,s)=e^{-\sqrt[+]{-2is}x}\hat\Phi(s).
\end{equation}
Then
\begin{equation}
\frac{\partial}{\partial
x}\hat\Psi_{2,3}(x,s)=-\sqrt[+]{-2is}\hat\Psi_{2,3}(x,s),\ \ \ \ x\ge 0.
\end{equation}
At the vertex ($x=0$) for bonds $B_{2,3}$ (``input'' from interior)
using the continuity boundary conditions \eqref{cont1} we get
\begin{equation}
\frac{\partial}{\partial
x}\hat\Psi_{2,3}(x=0,s)=-\sqrt[+]{-2is}\frac{\alpha_1}{\alpha_{2,3}}\hat\Psi_1(x=0,s).
\end{equation}
The Laplace transformed current conservation (at $x=0$) takes the
form
\begin{align}
\frac{\partial}{\partial
x}\hat\Psi_1&=\frac{\alpha_1}{\alpha_2}\frac{\partial}{\partial x}\hat\Psi_2 +
\frac{\alpha_1}{\alpha_3}\frac{\partial}{\partial x}\hat\Psi_3\nonumber\\
&=-\sqrt[+]{-2is}\
\alpha_1^2\left(\frac{1}{\alpha_2^2}+\frac{1}{\alpha_3^2}\right)\hat\Psi_1.
\end{align}
Using the inverse transform we have
\begin{equation}
\frac{\partial}{\partial
x}\Psi_1(x=0,t)=A_1\sqrt{\frac{2}{\pi}}e^{-i\frac{\pi}{4}}\frac{d}{d
t}\underset{0}{\overset{t}{\int}}{\frac{\Psi_1(0,\tau)}{\sqrt{t-\tau}}d\tau},\label{tbcgr}
\end{equation}
with $A_1=\alpha_1^2\left(\alpha_2^{-2}+\alpha_3^{-2}\right).$

The boundary condition given by \eqref{tbcgr} coincides with that in
Eq.~\eqref{tbc00} and thereby providing reflectionless transmission for the
bond $B_1$, when $A_1=1$, i.e.\ the following sum rule is fulfilled:
\begin{equation}
\frac{1}{\alpha_1^2}=\frac{1}{\alpha_2^2}+\frac{1}{\alpha_3^2}.\label{sumrule}
\end{equation}
Hence, the vertex boundary conditions given by Eqs.~\eqref{cont1}
and \eqref{Kirch1} become equivalent to the transparent vertex
boundary conditions, provided the sum rule in Eq.~\eqref{sumrule}
is fulfilled.

The numerical example in Fig.~\ref{fig2} shows a simulation of a right traveling Gaussian wave packet
\begin{equation*}
\Psi^I(x)=(2\pi)^{-1/4}\exp(5ix-(x+5)^2/4)
\end{equation*}
at four consecutive time steps. The Crank-Nicolson finite difference scheme
with the space discretization $\Delta x=0.016$ and the time step $\Delta
t=5\cdot10^{-5}$ has been used. It is clear from this figure that the wave
completely transmits to the second and third bonds when time elapses. We note
that the above boundary conditions provide conservation of the total norm,
which is defined as the sum of partial norms for each bond. Detailed derivation
of the norm conservation is provided in Appendix.

\begin{figure}[th!]
\includegraphics[width=90mm]{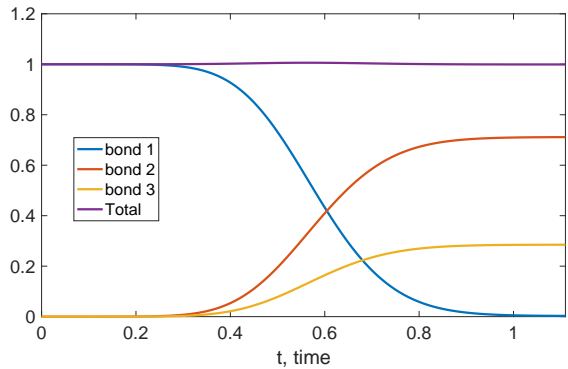}
\caption{(Color online). Time dependence of the partial and total norms for the
case shown in Fig.2.} \label{norm}
\end{figure}

\begin{figure}[th!]
\includegraphics[width=90mm]{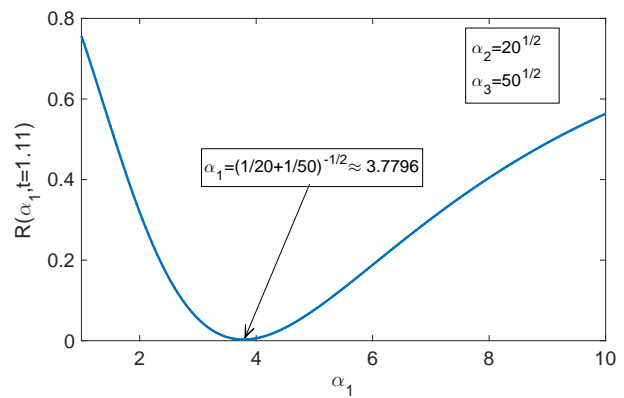}
\caption{Dependence of the vertex reflection coefficient $R$ on the parameter
$\alpha_1$ when time elapses ($t=1.1$).} \label{alpha}
\end{figure}

Fig.~\ref{norm} presents plots of the partial and total norms for this case. In
Fig.~\ref{alpha} the reflection coefficient determined as the ratio of the
partial norm for the first bond to the total norm
\begin{equation*}
R=\frac{N_1}{N_1+N_2+N_3}
\end{equation*}
is plotted as a function of $\alpha_1$ for the fixed values of $\alpha_2$ and
$\alpha_3$. As this plot shows, at the value of $\alpha_1$ that provides
fulfilling of the sum rule \eqref{sumrule} the reflection coefficient becomes
zero. This confirms once more that the sum rule in Eq.~\eqref{sumrule} makes
the vertex boundary conditions in Eqs.~\eqref{cont1} and \eqref{Kirch1}
equivalent to the transparent ones.

\begin{figure}[th!]
\includegraphics[width=80mm]{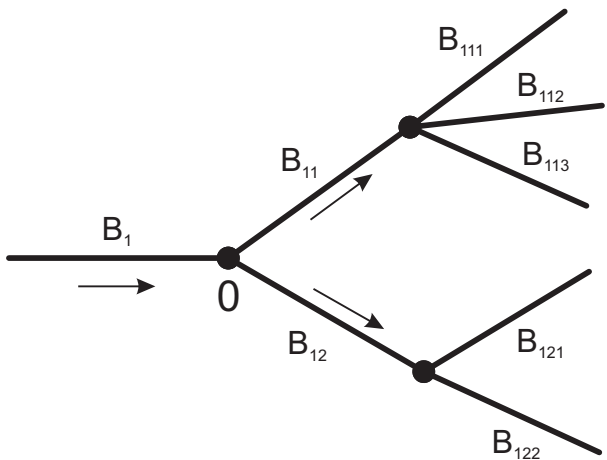}
\caption{Tree graph with three layers, $B_1{\sim}(-\infty, 0]$, $B_{11}$,
$B_{12} \sim [0,L_k)$, $k=1,2$, and $B_{1ij} \sim [0,+\infty)$ with
$i,j=1,2,3$.} \label{tree}
\end{figure}

We note that the above approach can be extended for other network branching
topologies, e.g., for tree graph presented in Fig.~\ref{tree}. The graph
consists of three ``layers'' $B_1$, $(B_{1i})$, $(B_{1ij})$, where $i,j$ run
over the given bonds. On each bond $B_1$, $B_{1i}$, $B_{1ij}$ we have the
time-dependent Schr\"odinger equation given by \eqref{lse0001}, for which
Kirchhoff-type boundary conditions are imposed at each vertex. Transparent
boundary conditions can be imposed similarly to those for the star graph. Then,
for all $i$, $j$, the $\alpha_{1i}$ and $\alpha_{1ij}$ have to be determined
from the sum rule like \eqref{sumrule} at each vertex. For instance, at the
three vertices in Fig.~\ref{tree} we have
\begin{equation}\label{r0}
\begin{array}{ll}
\text{ end of } B_1:\quad & \alpha_1^{-2}=\alpha_{11}^{-2}+\alpha_{12}^{-2},\\
\text{ end of } B_{11}:\quad & \alpha_{11}^{-2}=\alpha_{111}^{-2}+\alpha_{112}^{-2}+\alpha_{113}^{-2},\\
\text{ end of } B_{12}:\quad &
\alpha_{12}^{-2}=\alpha_{121}^{-2}+\alpha_{122}^{-2}.
\end{array}
\end{equation}
Fulfilling of these sum rules turns the Kirchhoff-type boundary conditions
equivalent to the transparent ones at the each vertex of a tree graph. The
constants $\alpha_j$ in Eq.~\eqref{sumrule} characterize the physical
properties of the material, which is used to construct the network and  can
have different meaning for each specific system under consideration.

%%%%%%%%%%%%%%%%%%%%%%%%%%%%%%%%%
\section{Conclusions}
In this paper we studied the problem of transparent quantum graphs by
determining them as branched quantum wires providing reflectionless
transmission of waves at the branching points. The boundary conditions for the
time-dependent Schr\"odinger equation on graphs, providing such transmission,
are formulated explicitly. A simple constraint that makes the usual
Kirchhoff-type boundary conditions at the vertex equivalent to those of
transparent ones is derived.

We have shown numerically for the star graph a reflectionless transmission of
the Gaussian wave packet through the vertices, if these constraints are
fulfilled. The approach can be extended straight forward for arbitrary graph
topologies, which contain any subgraph connected to two or more outgoing,
semi-infinite bonds. Stability analysis for the boundary conditions
discretization scheme is not included into the above treatment and will be
subject of forthcoming papers.

The motivation for the study of transparent quantum graphs comes from different
practically important problems of optics, condensed matter and polymers. One of
the problems allowing direct application the above model is charge carrier
transport in conducting polymers, which are highly branched nanoscale
structures. Reflectionless transmission of charge carriers in these structures
causes high conductivity and optimal functioning of different organic
electronic devices.
One of such structures was discussed recently in the
context of nonlinear charge carriers in branched conducting polymers \cite{Chsol}.
More attractive applications are possible also in linear fiber
networks, where the wave dynamics is described in terms of the Helmholtz equation.

Finally, we note that extension of the above approach of other linear
wave equations, such as heat, diffusion, Dirac and Klein-Gordon equations
should be rather straightforward.
Moreover, the method can be modified and
applied for the nonlinear partial differential equations widely used in
physics.

 %%%%%%%%%%%%%%%%%%%%%%%%%%%%%%%%%%%%%%%
\section{Acknowledgements}

This work is supported by the grant of the Ministry for Innovation
Development of Uzbekistan (Ref. Nr. BF2-022).
The work of JRY is
supported by Deutscher Akademischer Austauschdienst (DAAD).

\medskip
\appendix
\section{Norm conservation}\label{NormApp}
For quantum star graph with three bonds, the total norm is given as
\begin{equation}
N(t)=\sum\limits_{j=1}^3{\int\limits_{B_j}{|\Psi_j(x,t)|^2}dx},
\end{equation}
where $B_j$ is the domain of the bond $j$.

For Eq.~\eqref{lse0001} an easy calculation shows that the
derivative of the norm with respect to $t$ reads
\begin{equation}
\dot N(t) =
\sum\limits_{j=1}^3{\int\limits_{B_j}{\left[-\frac{i}{2}\frac{\partial^2\Psi_j^*}{\partial
x^2}\Psi_j+\frac{i}{2}\frac{\partial^2\Psi_j}{\partial
x^2}\Psi_j^*\right]dx}}
\end{equation}

Taking into account that $\Psi_1$ vanishes as $x\to-\infty$ and
$\Psi_{2,3}$ vanish as $x\to +\infty$ we obtain
\begin{align}
\dot N(t)&=\Im m\left[\Psi_1\frac{\partial\Psi_1^*}{\partial
x}\right]_{x=0}\nonumber\\
&-\Im m\left[\Psi_2\frac{\partial\Psi_2^*}{\partial x}\right]_{x=0}-\Im
m\left[\Psi_3\frac{\partial\Psi_3^*}{\partial x}\right]_{x=0},
\end{align}
and using boundary conditions \eqref{cont1} and \eqref{Kirch1}
we get
\begin{equation}
\dot N(t)=0,
\end{equation}
which means the preservation of the norm over time.

%%%%%%%%%%%%%%%%%%%%%%%%%%%%%%%%%%%%%%%%%%


\begin{thebibliography}{99}

\bibitem{BWG1} R. Wilson,  T.J. Karle,  I. Moerman, T.F. Krauss,  J. Opt. A: Pure Appl. Opt., \textbf{5} S76 (2003).
\bibitem{BWG2} P.I. Borel, , L.H. Frandsen,  A. Harp{\o}th,{\it et. al.}   Electron. Lett., \textbf{41} 69 (2005).
\bibitem{BWG3} S.V. Boriskina, Opt. Express, \textbf{15}  17371  (2007).
\bibitem{BWG4} W. Yang, X. Chen,  X. Shi,  W. Lu, Physica B, \textbf{405} 1832  (2010).
\bibitem{BWG5} J.B. Driscoll, R.R. Grote, B. Souhan, Opt. Lett.,  \textbf{38} 1854 (2013).
\bibitem{BWG6} G. Berkolaiko, R. Carlson, S.A. Fulling and P. Kuchment (eds.),  {\it Quantum Graphs and Their Applications},
Contemporary Mathematics 415, AMS, 2006.
% DOI: http://dx.doi.org/10.1090/conm/415
\bibitem{BWG7} P. Kuchment,
% Quantum graphs: an introduction and a brief survey
 arXiv preprint arXiv:0802.3442, 2008.

\bibitem{Arnold1998} A. Arnold and M. Ehrhardt, J. Comput. Phys., \textbf{145(2)}, 611-638 (1998).
\bibitem{Ehrhardt1999} M. Ehrhardt, VLSI Design, \textbf{9(4)}, 325-338 (1999).
\bibitem{Ehrhardt2001} M. Ehrhardt and A. Arnold,
Riv. di Math. Univ. di Parma, \textbf{6(4)}, 57-108 (2001).
\bibitem{Ehrhardt2002} M. Ehrhardt, Acta Acustica united with Acustica, \textbf{88}, 711-713 (2002).
\bibitem{Arnold2003} A. Arnold, M. Ehrhardt and I. Sofronov, Comm. Math. Sci., \textbf{1(3)}, 501-556 (2003).
\bibitem{Jiang2004} S. Jiang,  L. Greengard, Comput. Math. Appl., \textbf{47}, 955 (2004).
\bibitem{Antoine2008} X. Antoine, A. Arnold, C. Besse, M. Ehrhardt and A. Sch\"adle, Commun. Comput. Phys. \textbf{4(4)}, 729-796 (2008).
\bibitem{Ehrhardt2008} M. Ehrhardt, Appl. Numer. Math. \textbf{58(5)}, 660-673 (2008).
\bibitem{Sumichrast2009} L. \u{S}umichrast and M. Ehrhardt, J. Electr. Engineering \textbf{60(2)}, 301-312 (2009).
\bibitem{Antoine2009} X. Antoine et al., J. Comput. Phys., \textbf{228(2)}, 312-335 (2009).
\bibitem{Ehrhardt2010} M. Ehrhardt, Numer. Math.: Theo. Meth. Appl. \textbf{3(3)}, 295-337 (2010).
\bibitem{Klein2011} P. Klein, X. Antoine, C. Besse and M. Ehrhardt, Commun. Comput. Phys. \textbf{10(5)}, 1280-1304 (2011).
\bibitem{Arnold2012} A. Arnold, M. Ehrhardt, M. Schulte and I. Sofronov, Commun. Math. Sci. \textbf{10(3)}, 889-916 (2012).
\bibitem{Feshchenko2013} R.M. Feshchenko and A.V. Popov, Phys. Rev. E \textbf{88}, 053308 (2013).
\bibitem{Antoine2014} X. Antoine et al., J. Comput. Phys., \textbf{277}, 268-304 (2014).
\bibitem{Petrov2016} P. Petrov and M. Ehrhardt, J. Comput. Phys. \textbf{313}, 144-158 (2016).


\bibitem{Uzy1} T. Kottos and U. Smilansky, Ann. Phys., \textbf{76} 274 (1999).
\bibitem{Hul} O. Hul et al., Phys. Rev. E \textbf{69}, 056205 (2004).
\bibitem{Kuchment04} P. Kuchment,
Waves in Random Media, \textbf{14} S107 (2004).
\bibitem{Uzy2} S. Gnutzmann and U. Smilansky, Adv. Phys. \textbf{55} 527 (2006).
\bibitem{Exner15} P. Exner and H. Kovarik, {\it Quantum waveguides.} (Springer, 2015).


\bibitem{Zarif} Z. Sobirov, D. Matrasulov, K. Sabirov, S. Sawada, and K. Nakamura, Phys. Rev. E \textbf{81}, 066602  (2010).
\bibitem{zar2011} Z. Sobirov, D. Matrasulov, S. Sawada, and K. Nakamura, Phys. Rev. E \textbf{84}, 026609 (2011).
\bibitem{Karim2013} K.K. Sabirov, Z.A. Sobirov, D. Babajanov, and D.U. Matrasulov,  Phys. Lett. A, \textbf{377}, 860 (2013).
\bibitem{caputo14}  J.-G. Caputo, D. Dutykh, Phys. Rev. E \textbf{90}, 022912 (2014).
\bibitem{Our2015}  H. Uecker, D. Grieser, Z. Sobirov, D. Babajanov and D. Matrasulov, Phys. Rev. E \textbf{91}, 023209 (2015).
\bibitem{Adami16} R. Adami, C. Cacciapuoti,  D. Noja, J. Diff. Eq. \textbf{260}, 7397 (2016).
\bibitem{Our1} Z. Sobirov, D. Babajanov, D. Matrasulov, K. Nakamura, and H. Uecker, EPL \textbf{115} , 50002 (2016).
\bibitem{Adami17} R. Adami, E. Serra, P. Tilli, Commun. Math. Phys. \textbf{352}, 387 (2017).
\bibitem{dimarecent} A. Kairzhan, D.E. Pelinovsky,  J. Phys. A: Math. Theor. \textbf{51}, 095203 (2018).
\bibitem{Karim2018} K.K. Sabirov, S. Rakhmanov, D. Matrasulov and H. Susanto,  Phys. Lett. A, \textbf{382}, 1092 (2018).
\bibitem{KarimNLDE} K.K. Sabirov, D.B. Babajanov, D.U. Matrasulov and P.G. Kevrekidis, J. Phys. A: Math. Gen. \textbf{51}  435203 (2018).
\bibitem{Our2018} K.K. Sabirov, M. Akromov, Sh.R. Otajonov, D.U. Matrasulov, arXiv:1808.10751.

\bibitem{Shibata} T. Shibata, Phys. Rev. B \textbf{43(8)}, 6760 (1991).
\bibitem{Kuska} J.-P. Kuska, Phys. Rev. B \textbf{46(8)}, 5000 (1992).

\bibitem{Pauling} L. Pauling, J. Chem. Phys. \textbf{4} 673 (1936).
\bibitem{Rud} K. Ruedenberg and C.W. Scherr, J. Chem. Phys. \textbf{21}, 1565 (1953).
\bibitem{Alex} S. Alexander, Phys. Rev. B \textbf{27}, 1541 (1985).

\bibitem{Exner1} P. Exner, P. Seba, P. Stovicek, J. Phys. A: Math. Gen. \textbf{21} 4009 (1988).
\bibitem{Kost} V. Kostrykin and R. Schrader  J. Phys. A: Math. Gen. \textbf{32} 595 (1999).

\bibitem{Bolte} J. Bolte and J. Harrison, J. Phys. A: Math. Gen. \textbf{36} L433 (2003).
\bibitem{KarimBdG} K.K. Sabirov, J. Yusupov, D. Jumanazarov, D. Matrasulov,  Phys. Lett. A, \textbf{ 382}, 2856 (2018).

\bibitem{Uzy4} S. Gnutzmann, H. Schanz and U. Smilansky, Phys. Rev. Lett., \textbf{110}, 094101 (2013).
\bibitem{Mugnolo} D. Mugnolo. {\it Semigroup Methods for Evolution Equations on Networks}. Springer-Verlag, Berlin, (2014).
\bibitem{Grisha} G. Berkolaiko, P. Kuchment, {\it Introduction to Quantum Graphs, Mathematical Surveys and Monographs} AMS (2013).
\bibitem{yusupov2015} J. Yusupov, M. Dolgushev, A. Blumen and
O. M\"ulken. Quantum Inf Process \textbf{15}, 1765 (2016).
\bibitem{Ourhe} K. Sabirov, Zh. Zhunussova, D. Babajanov, D. Matrasulov, arXiv:1806.10957.
\bibitem{Chsol} D. Babajanov, H. Matyoqubov, D. Matrasulov, J. Chem. Phys.,  \textbf{149}, 164908 (2018).

\end{thebibliography}
\end{document}